# Carbon nanotubes for the optical far-field readout of processes that are mediated by plasmonic near-fields.

M. Glaeske[1], P. Kusch[1], N. S. Mueller[1,$], and A. Setaro[1,2,*]

[1]Department of Physics, Freie Universität Berlin, 14195 Berlin, Germany.

[2]Pegaso University, Naples, 80132, Italy.

[$]Present address: NanoPhotonics Centre, Cavendish Laboratory, University of Cambridge, Cambridge, CB3 0HE, United Kingdom.

*Corresponding author: setaro@physik.fu-berlin.de , antonio.setaro@unipegaso.it

**Abstract**

As science progresses at the nanoscopic level, it becomes more and more important to comprehend the interactions taking place at the nanoscale, where optical near-fields play a key role. Their phenomenology differs significantly from the propagative light we experience at the macroscopic level. This is particularly important in applications such as surface-enhanced spectroscopies for single-molecule detection, where often the optimization of the plasmonic structures and surfaces relies on far-field characterizations. The processes dominating in the far-field picture, though, are not the same dominating in the near-field. To highlight this, we resort to very simple metallic systems: Isolated gold nanorods in solution. We show how single-walled nanotubes can be exploited to read out processes occurring at the near-field level around metallic nanoparticles and make the information accessible in the far-field region. This is implemented by monitoring the spectral profile of the enhancement of the photoluminescence and Raman signal of the nanotubes for several excitation wavelengths. Through this excitation-resolved study, we show that the far-field optical read-out detects the transversal and longitudinal dipolar plasmonic oscillations of gold nanorods, whereas the near-field read-out through the nanotubes reveals other mechanisms to dominate. The spectral position of the maximum enhancement of the optical near-field mediated signals are located elsewhere than the far-field bands. This dichotomy between near-field and far-field response should be taken into account when optimizing plasmonic nanostructures for applications such as surface-enhanced spectroscopies.

**Introduction**

When a light beam hits an object of finite size, it is always converted into both a propagating and an evanescent field [1]. The propagative part, described as superposition of plane waves [2],[3], can freely travel through space to the so-called far-field region. The evanescent fields fade out very rapidly and remain localized around the surface of the object, whose immediate surrounding is referred as near-field region [2],[3]. While propagating far-fields possess a dipolar character, near-fields can exhibit multipolar character [3],[4]. Dark modes, which are not directly optically detectable by far field techniques, also play an important role in the near-fields, for example for transferring energy [5]. The near-field surrounding an object can reveal precious information about the object itself. Microscopy based on near-fields, for example, has allowed to break the intrinsic boundaries of optical spectroscopy dictated by Abbe's diffraction limit, enabling the detection of nanometre scale objects much smaller than the wavelength of visible light [6].

Metallic nanostructures can efficiently confine light to the nanoscale [7],[8]. The collective motion of their free conduction electrons yields a large oscillating dipole moment that efficiently couples with radiation in the visible spectral range [7] and leads to the localized surface plasmons [8]. The oscillating plasmonic dipole generates an intense near-field close to the metal surface. This enables the concentration of light in nanoscale volumes, which can be exploited for *e.g.* increasing the efficiencies of photodetectors, solar energy collectors, and catalysis [9],[10]. The plasmonic near-fields of structured metallic surfaces have pushed the molecular recognition to the single molecule level [11],[12], enhancing the photoluminescence and Raman yield of nanomaterials placed in their close proximity [13]-[15]. The enhancement of the Raman scattering cross section can reach ten orders of magnitude, which is known as surface-enhanced Raman scattering (SERS) [16]. The strongly localized near-fields efficiently couple with the elementary excitations of molecules and nanomaterials, allowing for the formation of hybridized polaritonic states through strong light-matter coupling [17]-[19]. Near-fields can locally overrule the flow of energy between donors and emitters to allow transfers that are usually forbidden through far-field radiative coupling [20],[21]. They have moreover attracted the interest of the scientific community for their ability to control multiphoton dynamics in multiparticle quantum systems [22].

These and further developments of near-field based science and applications rely on the careful near-field design with *ad hoc* tailored features. Unfortunately, this is not straight forward. The design of the substrates for enhanced spectroscopy based on the far-field response of the plasmonic nanostructure might lead to wrong expectations about their performances, since the mechanisms that are most efficiently observed at the far-field might not occur at the same wavelength as near-field mediated processes. The energies/wavelengths obtained by far-field means are not the ones that automatically ensure the highest emission yields. The experimental read-out of the near-fields by all-optical means is a challenging task that, for the very same elusive nature of the near-fields, cannot be performed with standard techniques, which rely on propagative fields. To achieve near-field detection through far-fields, the near-fields need to be converted from a localized to a propagative field. The most advanced and reliable technique developed so far for this purpose is scanning near-field optical microscopy (SNOM), where near-fields are detected through tiny apertures and very sharp tips [6],[23],[24]. Those advanced setups require specialized users and stable, controlled lab conditions (temperature, air flow, vibrations, and so on), making SNOM an advanced but yet not widespread technique for the direct optical readout of the near-field features of a desired object or surface. Moreover, the typical commercial SNOM setups have only a single excitation wavelength, which prevents to perform resonant studies. Worldwide there are only very few special SNOMs setups able to perform wavelength-resolved studies. Excitation-wavelength resolved SERS investigations of molecules and nanomaterials interfaced with specialized plasmonic nanostructures have been successfully performed in an alternative approach [25]-[28].

In this work we suggest to use the emission (photoluminescence or Raman) of single-walled carbon nanotubes (SWNTs) for the far-field optical readout of the processes that occur in the presence of plasmonic near-fields. Studies performed with molecular emitters have successfully demonstrated optical enhancement and, by taking advantage of the small size of the emitters with respect to the size of the metallic structures, the capability of mapping the field localization around the metallic particles [29]. The advantage we reckon for carbon nanotubes is the spectral extension of their excitation window, which covers the entire spectral range achievable by colloidal gold. Moreover, nanotubes have a big separation between the emission and excitation windows, whereas the Stokes shift in molecular emitters is much smaller. The photoluminescence background in the SERS spectra is therefore much lower for SWNTs. There are plenty of different SWNTs species, each described uniquely by their chiral vector $\vec{c} = (n_1, n_2)$, with $n_1$ and $n_2$ integers, and its distinctive excitation ($E_{22}$) and emission ($E_{11}$) transition energies [30]. Those transition energies span across a broad spectral range [30],[31], making carbon nanotubes ideal candidates for the optical read-out of near-fields from the visible up to the near-infrared. SNOM measurements of aligned SWNTs deposited onto optical Hertzian dipole antennas proved the capability of SWNTs as near-field detectors [32]. We use the SWNTs to read out the optical response from near fields specific processes of gold nanorods in solution. With wavelength-scanned PL and SERS we demonstrate that even for this simple plasmonic system, there is a marked difference between both read-outs.

While the far-field reading is dictated by the plasmonic response of the gold nanorods (strong transversal and longitudinal dipolar oscillations), the one mediated by the near-fields is strongly influenced by their coupling with the elementary excitations of the material. As a result, we observe a spectral shift between the far-field resonances and the wavelengths where the plasmon-enhanced photoluminescence and Raman signals are the strongest. This spectral shift has to be taken into account when designing plasmonic nanostructures for maximum enhancement at a specific target wavelength.

**Experimental details and methods**

The SWNTs used in this work were purchased from SouthWest NanoTechnologies (SWeNTs) and produced according to the CoMoCATs® method, batch SG76, with diameter between 0.8 and 1.2 nm and median length of 300 nm. All the other chemicals were purchased from Sigma–Aldrich.

We produced our gold nanorods (AuNRs) via the seed mediated growth method described in Refs. [33] and [34]. The procedure requires a seed and a growth solution to be prepared independently and then mixed together. We accordingly prepared first a stock solution of 5 mL distilled water, 0.1 M CTAB and 0.5 mM of $HAuCl_4$. We then heated up this stock solution to 25–30°C until the CTAB was fully dissolved and then added 0.3 mL of $NaBH_4$ (0.01 M), which made the solution turn light brown. The last step to finalize the seed solution required the decomposition of $NaBH_4$, achieved by heating up the stock solution up to 40–45°C for 15 min. In addition, we prepared the growth solution with 10 mL distilled water, CTAB (0.1 M) and $HAuCl_4$ (0.5 mM). We then added 0.75 mL of ascorbic acid to the growth solution, 90 mL of silver nitrate to drive the anisotropic growth of the rods and finally 20 mL of the seed solution to trigger the growth. After 10 minutes of gentle stirring, the solution turned purple. We stopped the stirring after 30 min and left the solution to settle. After three days, the CTAB surplus sedimented and we could use the AuNRs supernatant for our experiments.

To place the SWNTs in the close proximity of the AuNRs, we followed the routine that we have developed and studied in the past to yield the metal-nanotubes hybrids, that we called π-hybrids [15],[35] to highlight the leading role of the π-conjugation of the carbon network into the optical activity of the nanotubes. Instead of following the standard isolation-by-sonication procedure [36],[37], the π-hybrids are prepared following a milder path: In an adaptation of the micelle-swelling technique [38],[39], we start with the solution of the AuNRs isolated in the CTAB micelles, add the pristine SWNTs and let them stir gently over two weeks. At the end of the stirring phase, a fraction of isolated tubes will have entered the micelle and lie in close proximity to the AuNRs; the other excess nanotubes will precipitate once the stirring is stopped. By following this procedure, we ensure that the nanotubes penetrate the micelles without destroying them. The micellar forces ensure good alignment between tubes and rods. Fig. 1a schematically sketches our π-hybrids in suspensions. To prepare the π-hybrids for this work, we added 210 μL of SWNTs in water (0.1 gL$^{-1}$, no additional surfactant) to 1 mL of the AuNR solution and left it stirring for two weeks. This concentration has been chosen to work in the low-filling regime of the micelles to avoid tube-tube interactions and make each nanotube sense only the plasmonic fields of the nanorods. We also prepared a reference sample consisting of 210 μL SWNTs solution stirred in 1 mL of a CTAB (0.1 M) solution.

The absorption spectra of our samples were recorded with a Perkin-Elmer Lambda 950 spectrophotometer. The photoluminescence of our samples has been obtained by a Nanolog spectrofluorometer from Horiba equipped with a Xenon arc lamp pump and a nitrogen-cooled InGaAs array detector. The Raman spectra have been acquired with a Horiba T6400K triple grating monochromator coupled with a CCD detector. As excitation source for the wavelength-dependent Raman experiment we used a Ti:sapphire (Coherent; 690 nm to 1050 nm), and a dye (Radiant dyes; R6G, DCM dyes; 570 nm to 680 nm) laser. The laser was focused by a 100x objective (N.A.= 0.9), which was also used to collect the backscattered light. To prevent damage of the samples by laser heating, we kept the laser power down to 1 mW. Each spectrum was acquired with an integration time of 120 s. The Raman spectra of pristine SWNTs and π-hybrids were acquired as a function of the excitation wavelength. With each measurement, we also recorded in the same scattering geometry a reference Raman spectrum of diamond, whose Raman cross section is constant. The diamond spectra (Raman mode at 1330 cm$^{-1}$) were applied to calibrate the intensity of the G band intensity. We divided the G band Raman intensity of the SWNTs by that of the 1330 cm$^{-1}$ diamond mode to correct for the wavelength-dependent sensitivity of the Raman spectrometer.

We simulated the optical response of the AuNRs with finite-difference time-domain (FDTD) simulations, using the commercial software package Lumerical FDTD Solutions. The gold nanorods were modelled as cylinders with rounded edges and assigned the optical dielectric function of gold. The dielectric function was obtained from a fit of experimental data from Johnson and Christy [40]. The gold nanorod was placed in a 1 μm$^3$ FDTD simulation cell with perfectly matched layers as boundaries. Space around the nanorod was discretized into 0.5 nm cells using a mesh-override region. A total-field scattered-field plane wave source was used to inject a 3 fs light pulse. The optical absorption was recorded with a box-like power monitor around the nanorod and the electric field with line monitors positioned 1 nm away from metal surface. The electric fields were recorded till the simulation converged and the wavelength dependence was obtained through a Fourier transformation. SERS enhancement factors were obtained by calculating the local SERS enhancement (see Eqs. 1 and 2 below) from the local electric field enhancement at each position of the line monitors. Average SERS enhancement factors were then obtained as the spatial average of the local SERS enhancement. We furthermore averaged over three excitation configurations: i) light propagation perpendicular to the nanorod axis and polarization along the nanorod axis, ii) the same but polarization perpendicular to the nanorod axis and iii) light propagation along the nanorod axis.

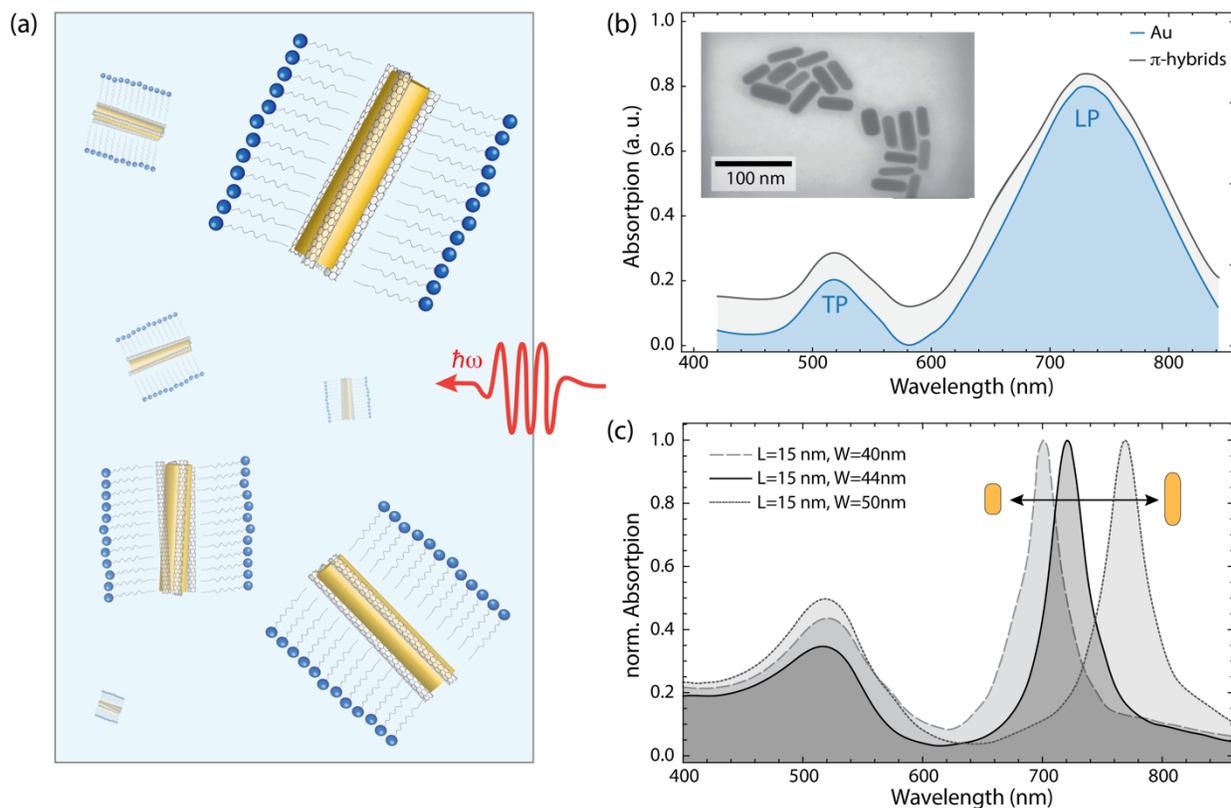

**Figure 1**: (a) Sketch schematically depicting an ensemble of π-hybrids in solution. (b) Representative experimental absorption spectrum of the AuNRs synthesized for this work and of the resulting π-hybrids. Inset: SEM image of the AuNRs dropcasted on a TEM grid. (c) FDTD simulation of the absorption spectra of AuNRs with different aspect ratios with length $L$ and width $W$. The spectra are averaged over three excitation configurations (see Methods).

**Results and discussion**

Our wet-chemistry synthesis approach yielded an ensemble of AuNRs in solution whose optical absorbance spectra exhibited the typical features of gold nanorods, with a transversal plasmonic (TP) resonance band peaked at 520 nm and a longitudinal plasmonic (LP) band peaked at 730 nm (Fig. 1b). This ensemble displays a certain polydispersity in the particle size (see the inset of Fig. 1b), which is typical for synthesis methods involving a wet-growth process [41]-[43]. Simulated absorption spectra are shown in Fig. 1c for selected aspect ratios and sizes of the AuNRs. The spectra were obtained from FDTD simulations, in which we averaged over three excitation configurations to excite both the TP and LP resonances (see Methods). The longitudinal and transverse plasmonic modes are well reproduced by the simulated spectra. The spectral width of the longitudinal mode is much narrower in the simulated spectra and its resonance wavelength depends on the aspect ratio of the AuNR. We attribute the broader spectral width in the measured spectrum to overlapping spectra from AuNRs with different aspect ratios and sizes. This is consistent with the literature observations of the broad absorption bands of unsorted samples containing a variety of particles with different shapes and aspect ratios [41]-[43].

The distinctive feature of the π-hybrids, and the reason why they were originally created for, is the enhancement of the SWNTs photoluminescence emission once they are placed in the close proximity of the AuNRs [15]. The modification of the optical response of the SWNTs by the metal surface can differ from system to system, ranging from enhancement to complete quenching. The π-hybrids operate in the so-called lossy regime, set by the SWNTs' low emission yields and the weak illumination conditions [35],[44],[45]. In such a regime, the maximum enhancement $\Phi_{PL}$ is achieved when the emitter is placed very close to the metal surface and quenching plays no significant role [44],[45]. In general, the enhancement arises from increased pumping $F_{exc}$ (due to the plasmonic near-fields) as well as increased radiative emission $F_{rad}$ (due to the additional metal-induced de-excitation channels) [13]: $\Phi_{PL} = F_{exc} F_{rad}$. For the π-hybrids $F_{exc}$ is the dominant enhancement mechanism [35]. Another noteworthy consequence of our operation conditions is that, in the lossy regime, the enhancement of first-order Raman processes behaves alike the photoluminescence enhancement [45]: $\Phi_{Raman} = F_{exc} F_{rad}$. In this case it is expected that the enhancement of photoluminescence and Raman follows a qualitatively similar wavelength dependence [45]. In the following, we will focus on the spectral response of the enhancement of the π-hybrids, discuss how it provides insights on the dominating components of the near-field enhancement, and how it relates to the far-field spectra.

We monitored the intensity of the Raman G band (which is a first-order process originating from the in-plane C-C vibrations [25]) and the PL emission of the tubes for different excitation wavelengths. Fig. 2a shows the PL emission spectra for different excitation wavelengths of the different pristine nanotube species present in our samples (in gray) compared with the emission from the π-hybrids (in blue). We chose excitation wavelengths (see labels) that are in resonance with the optical $E_{22}$ transitions (absorption window) of the different SWNT species present in our sample. Fig. 2b shows the Raman spectra of the G band (excited at the same wavelength used for the PL emission spectra) of the pristine nanotubes (in gray) compared with those of the π-hybrids (in red). Both, the PL and Raman spectra are enhanced by the presence of the AuNRs by ~250%. Fig. 2c shows the photoluminescence (in blue) and Raman enhancement factors (in red) for the different excitation wavelengths. They display a consistent and similar trend with a maximum of the enhancement peaked around 670 nm.

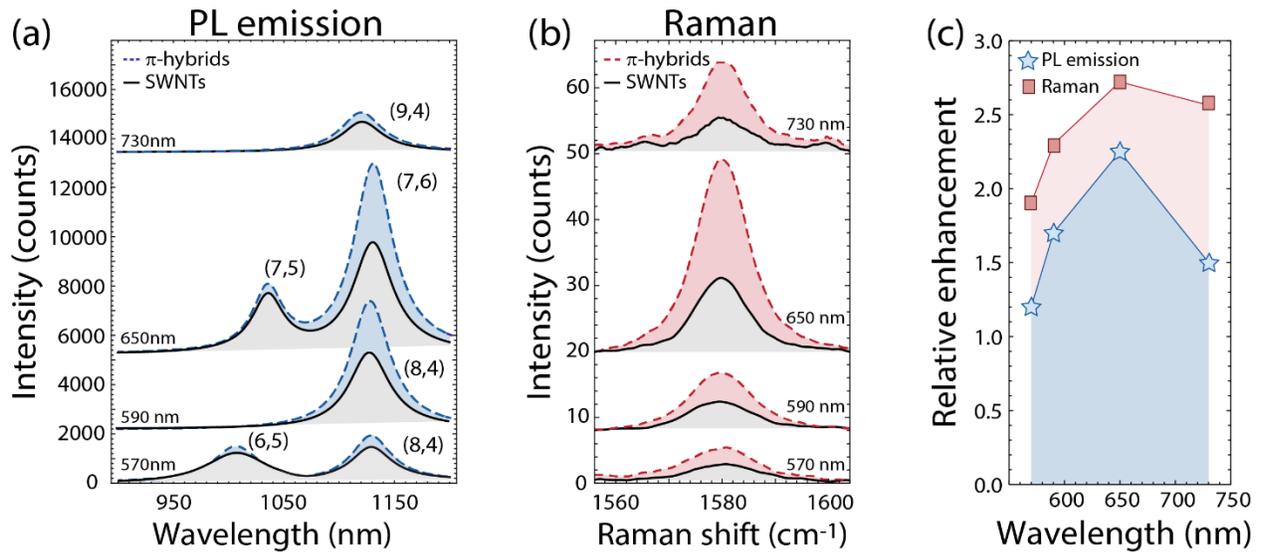

**Figure 2:** a) Comparison of the emission spectra of the π-hybrids (blue, dashed) and pristine nanotubes (gray) for different excitation wavelengths. Labels ($n_1,n_2$) indicate the chiral species of the SWNTs excited at the specific wavelength. The spectra are offset for better visibility. b) Comparison of the Raman G band of the π-hybrids (red, dashed) and pristine nanotubes (gray) for different excitation wavelengths. The spectra are offset for better visibility. c) Comparison of the spectral distribution of photoluminescence (blue, stars) and the Raman G band (red, squares) enhancement.

Both photoluminescence and Raman optical signals show the same enhancement profile, which is mediated by the plasmonic near-field. It is worth noting at this point the striking difference between the optical read-out of the plasmonic response by the far-field based UV/Vis technique, which is able to read-out the TP and LP band, peaked respectively at 520 nm and 730 nm (Fig. 1a), and the near-field readout through the enhancement of the emission and the Raman response, peaked at 670 nm.

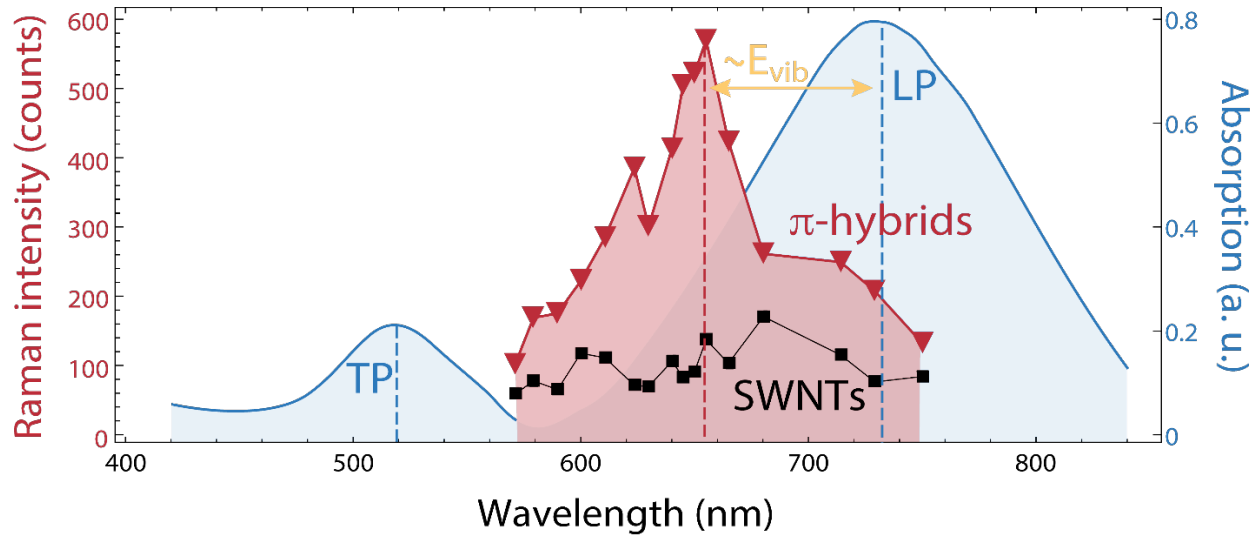

**Figure 3:** Intensity of the Raman G band of the π-hybrids (red triangles) compared with the one of the pristine tubes (black squares). The blue curve shows for comparison the absorption spectrum of the AuNRs.

To better resolve the enhancement profile, we performed resonant Raman studies of the optical response of our hybrids. Fig. 3 compares the intensity of the Raman band of the pristine nanotubes sample (black squares) with the one of the π-hybrids (red triangles). While the Raman intensity for the pristine nanotubes is wavelength independent, the intensity of the π-hybrids' Raman signal shows a marked increase around 650 nm. The origin of this enhancement, as discussed before, relies in the pumping by the near-field localized around the surface of the metallic particle. By comparing the resonance profile with the absorption spectrum, it clearly does not correspond to either the TP or the LP band of the AuNRs (blue curve). In the following, we will resolve this discrepancy which is at the heart of near-field detection by the SWNTs.

Near field resonances are usually red-shifted with respect to the far-field in localized plasmonic systems [46]. A straightforward way of explaining this effect has been elaborated by Nordlander and coworkers by depicting the plasmon oscillation as a driven, damped harmonic oscillator [46]. The features observed in the far-field, such as scattering or absorption, are related to the ability of the impinging radiation to accelerate and drive the conduction electrons into oscillations. The physics of the damped, driven harmonic oscillator reveals that this process is most efficient when the frequency of the driving force (*i.e.* the incoming radiation) matches the intrinsic frequency of the oscillator (*i.e.* the plasmon frequency $\omega_\pi$) [46]. That is the reason why the far-field excitation and detection work most efficiently at the plasmon frequency. The amplitude of the near-field surrounding the particle, on the contrary, follows the potential energy, which is maximized at a smaller frequency than the intrinsic one due damping effects of the oscillator [46]. Since damping is unavoidable, the near-fields surrounding small metallic particles are always maximized at smaller frequencies/longer wavelengths than the ones one would obtain by far-field readouts. To test whether this red-shift would account for the discrepancy between the TP maximum of the absorption spectra and the one of our enhancement profile, we calculated, by means of the FDTD simulations, the amplitude of the electric field surrounding a representative gold nanorod (width of 15 nm and length of 44 nm) for different excitation wavelengths (Fig. 4).

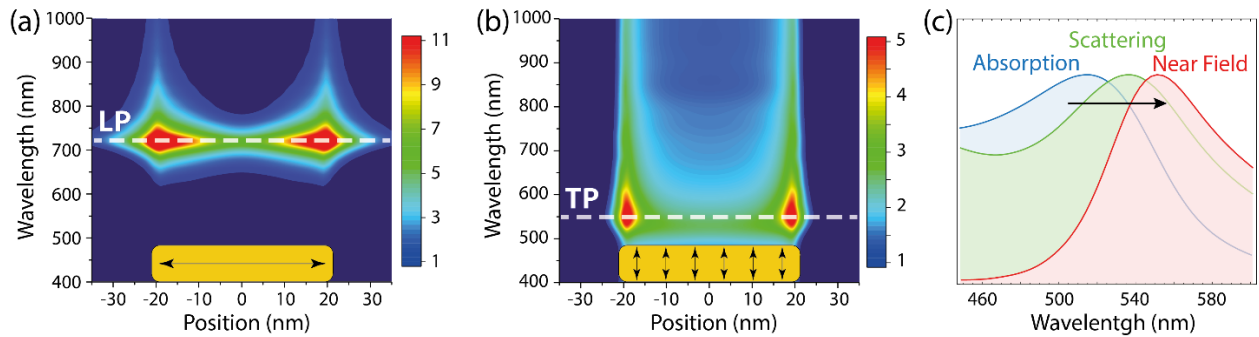

**Figure 4:** a) Amplitude of the electric field surrounding a gold nanorod excited with linearly polarized light parallel to the long axis of the rod to activate the longitudinal plasmonic mode (LP). b) Amplitude of the electric field surrounding a gold nanorod excited with linearly polarized light parallel to the short axis of the rod to activate the transversal plasmonic mode (TP). The near fields were recorded with a line monitor parallel to the long axis of the nanorod that was positioned 1 nm away from the nanorod surface and continued in the same direction past the nanorod edges. c) Comparison between normalized absorption (blue), scattering cross section (green), and near-field amplitude (red) for the TP mode, excited with light polarized along the short axis of the rod. The gold nanorod had a width of 15 nm and length of 44 nm, which is a typical size estimated from TEM images.

Fig. 4a reports the dependence of the electric field amplitude surrounding the metallic nanoparticle when irradiated with linearly polarized light parallel to its long axis to excite the LP oscillation mode. Fig. 4b reports the dependence of the electric field amplitude surrounding the metallic nanoparticle when irradiated with linearly polarized light parallel to its short axis to excite the more energetic TP oscillation mode. Fig. 4c compares the normalized absorption (blue) with the scattering cross section (green) and near-field amplitude (red) for the TP mode. Those data confirm the red-shift between the far-field and the near-field resonances, but its amount is too small to explain the SERS enhancement profile by the near field resonance of TP mode. Bigger near-field/far-field shifts could be obtained with tighter optical confinement, which does not occur in our system as nanorods aggregation is prevented by the micelles. Monitoring the intensity of the electric field localized near the AuNR surface rules also out the contribution of the near-field generated by the TP plasmon: The electric field generated by the TP oscillation is less intense than the one generated by the LP (cfr. the scale bars in Fig. 4a,b), making the LP oscillation the most efficient enhancement channel. The maximum of the enhancement profile is thus blue-shifted with respect to the dominant plasmon resonance, hinting at a different underlying enhancement mechanism.

Commonly, in the SERS framework, the enhancement factor is assumed to scale as the fourth power of the intensity of the electric field [16],[47]:

$$\Phi_{SERS}^{E^4} = \frac{1}{S} \int E^4(\omega_L) dS. \quad (1)$$

This result is obtained by neglecting the Raman shift ($E_{vib} = \hbar\omega_{vib} = 0$), an approximation that is often not justified [47]. A more accurate description of the SERS enhancement factor is given by [16],[47],[48]:

$$\Phi_{SERS} = \frac{1}{S}\int E^2(\omega_L)E^2(\omega_L - \omega_{vib})dS. \qquad (2)$$

The energy of the vibrational mode for the G band of SWNTs is $E_{vib} = \hbar\omega_{vib} = 196$ meV. Fig. 5a shows the calculated distribution of SERS enhancement $\Phi_{SERS}$ at the surface of the AuNR at different wavelengths of incoming radiation, which is linearly polarized along the long axis of the nanorod. One resonance band occurs at around 730 nm, where the electric field is maximally amplified by the LP oscillation (cfr. Fig. 4a). A second band arises at ~650 nm, which stems from an outgoing resonance (OR) in the SERS process, when the frequency of the Raman-scattered light $\omega_L - \omega_{vib}$ matches the LP plasmon. This is the same spectral window in which we observe the strongest SERS enhancement in our experimental data.

Fig. 5b compares the surface-averaged enhancement factors calculated as the $E^4$ (red, dashed curve) vs. the SERS enhancement factor (blue curve) for light linearly polarized along the long axis of the nanorod. The $E^4$ enhancement $\Phi_{SERS}^{E^4}$ follows the wavelength dependence of the electric near field (compare to Fig. 4a), which is strongest at the LP resonance and cannot account for our experimentally observed enhancement profile. The SERS enhancement factor $\Phi_{SERS}$, on the contrary, displays one additional band, due to the outgoing resonance (labelled as OR in the figure), which lies in the proper wavelength range to account for our experimental observations. Despite the appearance of the OR band, there is still a discrepancy between the calculated SERS enhancement profile $\Phi_{SERS}$, and our data: The peak enhancement of the calculated $\Phi_{SERS}$ profile resides at the LP position, whereas the experimental enhancement is largest at the wavelength of the outgoing resonance (cfr. Fig. 3).

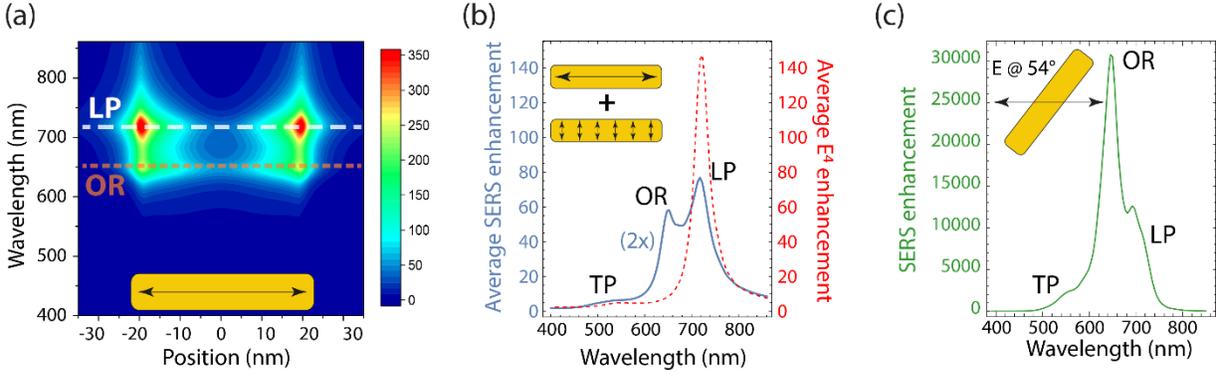

**Figure 5:** a) Map of the locally resolved SERS enhancement factor for different excitation wavelengths. The incident light was linearly polarized along the long axis of the nanorod, exciting the longitudinal plasmonic mode (LP). b) Comparison between the surface-averaged SERS enhancement factor $\Phi_{SERS}$ (blue curve, magnified by a factor of two for better visibility) calculated according to eq. 2 and the surface-averaged enhancement factor $\Phi_{SERS}^{E^4}$ calculated according to eq. 1 (red dashed curve) averaging over three different excitation configurations: 1. light incident perpendicular to the rod axis and polarization along the rod axis; 2. light incident perpendicular to the rod axis and polarization perpendicular to the rod axis; 3. light incident along the nanorod axis. c) SERS enhancement factor $\Phi_{SERS}$ evaluated at the uppermost corner of the AuNR for light polarized along the nanorod axis, impinging at an angle of 54° with respect to the nanorod axis. The AuNRs had a width of 15 nm and length of 44 nm – as in Fig. 4.

To explain this last discrepancy, we need to consider that the excitation processes devised in our calculations describe AuNRs illuminated with light linearly polarized along either the short or the long axis of the rod. In these configurations either the TP or LP plasmonic mode is excited selectively. Our π-hybrids sample consists of an ensemble of randomly isotropically oriented nanorods in solution, as sketched in Fig 1a. In these conditions, the illumination process does not ensure the pump light to be polarized strictly parallel to the long or the short axis of the AuNRs but rather at skew angles. Fig. 5c shows the SERS enhancement profile calculated at the uppermost corner of an AuNR for linearly polarized light impinging at an angle of 54° (not along any high symmetry direction) with respect to the long axis of the rod. This configuration, better fitting the real experimental conditions of our samples, excellently describes our experimentally observed resonance profile. The reason for the OR band becoming dominant in this configuration is that, at skew angles, both TP and LP plasmon resonances are excited. Their spectral overlap leads to an asymmetry of the near field enhancement $E^2(\omega_L)$, which, when multiplied as $E^2(\omega_L)E^2(\omega_L - \omega_{vib})$ leads to the stronger outgoing resonance OR band.

As our measurements were performed in solution, we probed an ensemble of randomly oriented nanorods. Most excitation configurations therefore lead to overlapping LP and TP resonances in the near field enhancement, which results in an overall dominating outgoing resonance in the experimental enhancement profile. This outgoing resonance is blue shifted with respect to the maximum near field enhancement by the Raman shift, which is in our case $E_{vib} = \hbar\omega_{vib} = 196$ meV. We envision a similar interplay of enhancement channels as the reason for the blue shifted PL emission, i.e. here the enhancement of absorption and emission. To perform more systematic wavelength-resolved studies of the photoluminescence enhancement profile, we plan future investigations with a richer set of SWNTs chiral species to cover a broader spectral range and gain higher resolution.

**Conclusion**

The characterization of metallic nanostructures solely with far-field optical tools can provide an incomplete picture of their performance in enhanced spectroscopy. The plasmonic resonances that appear as peaks in the far-field spectra can occur at wavelengths that are very different from the peak enhancement at the near-field level. This is the case we have shown in this manuscript, highlighting the dichotomy of the behavior of metallic nanorods: At the far-field level of observation, we observe their very strong transversal and longitudinal dipolar oscillations, which efficiently couple with propagative light and are the only ones self-selectively observed in the far field. The enhancement profile of photoluminescence and Raman scattering does not follow the far-field profile, showing that near-field-dictated processes behave differently. Monitoring the emission intensity of carbon nanotubes placed in the close proximity of metallic nanoparticles highlights their potential as sensors of mechanisms occurring at the near-field level, which are otherwise not directly observable with far-field techniques, such as optical absorption spectroscopy. Resonant Raman studies performed on colloidal gold nanorods-carbon nanotubes hybrids solutions prove the outgoing resonance of the SERS process to be dominating and neither occurring at the wavelength of the longitudinal nor of the transversal plasmons. These results confirm the ability of SWNTs to convert information about near-field related processes into propagative light, making them valid candidates for far-field optical readout of the elusive near-fields. We believe that this work can serve as a stimulus for pursuing alternative, simple protocols for reading out the elusive interaction mechanisms involving optical near-fields and optimize the plasmonic systems to be exploited for surface-enhanced spectroscopies.

**Acknowledgements**

This work was supported by the Focus Area NanoScale at Freie Universität Berlin. N.S.M. acknowledges support from the Deutsche Telekom Stiftung and the German National Academy of Sciences Leopoldina. The authors gratefully thank S. Reich for her valuable support and S. Jürgenssen for the scientific discussions.